\documentclass[twocolumn]{aa}
\usepackage{graphicx}
\usepackage{txfonts}
\usepackage{epstopdf}
\usepackage[authoryear]{natbib}
\bibpunct{(}{)}{,}{a}{}{;}
\begin{document}
\title{Structure of the Large Magellanic Cloud from Near Infrared magnitudes of Red clump stars}
\author{Smitha Subramanian\inst{1,2}, Annapurni Subramaniam\inst{1}}
\institute{Indian Institute of Astrophysics, Koramangala II Block, Bangalore-560034, India\\
	   Department of Physics, Calicut University, Calicut, Kerala\\
           \email{smitha@iiap.res.in, purni@iiap.res.in}}
\date{Received, accepted}
\abstract
{The structural parameters of the disk of the Large Magellanic Cloud (LMC) are estimated.}
{We used the {{\it JH}} photometric data of red clump stars  
from the Magellanic Cloud Point Source Catalog (MCPSC) obtained from the InfraRed Survey Facility (IRSF) 
to estimate the structural parameters of the
 LMC disk, like the inclination, $i$ and the position angle of the
 line of nodes (PA$_{lon}$), $\phi$.} 
 {The observed LMC region is divided into several sub-regions and stars in each region are 
cross identified 
with the optically identified red clump stars to obtain the near infrared magnitudes. 
The peak values of H magnitude and (J$-$H) colour of the observed red clump distribution 
are obtained by fitting a profile to the distributions and also by taking the average value of 
magnitude and colour of the red clump stars in the bin with largest number.
Then the dereddened peak H$_0$ magnitude of the red clump stars in each sub-region 
is obtained from the peak values of H magnitude and (J$-$H) colour of the observed red clump 
distribution. The RA, Dec and relative distance from the center of each sub-region are converted into x, y \& z Cartesian 
coordinates. A weighted least square plane fitting method is applied to 
this x,y,z data to estimate the structural parameters of the LMC disk.} 
{An intrinsic $(J$-$H)_0$ colour of 0.40 $\pm$ 0.03 mag in the IRSF SIRIUS filter system is estimated 
for the RC stars in the LMC and a reddening map based on (J$-$H) colour of the RC stars is presented. When the peaks of the red clump distribution were identified by averaging, an inclination of 
25$^o$.7$\pm$1$^o$.6 and PA$_{lon}$ = 141$^o$.5$\pm$4$^o$.5 were obtained. We estimate a distance modulus, 
$\mu$=18.47$\pm$0.1 mag to the LMC. Extra-planar features 
which are in front as well as behind the fitted plane are identified which match with the 
optically identified extra-planar features. The bar of the LMC is found to be part of the disk within 500 pc.}
{The estimates of the structural parameters are found to be independent of the photometric bands used for the 
analysis. The radial variation of the 
structural parameters are also studied. We find that the inner disk, within $\sim$ 3$^o$.0, is less 
inclined and has larger value of PA$_{lon}$ when compared to the outer disk. Our estimates are compared 
with the literature values and the possible reasons for the small discrepancies found are discussed.
}

\authorrunning{Subramanian \& Subramaniam}
\titlerunning{Structure of the Large Magellanic Cloud}
\maketitle

\section{Introduction}
The Large Magellanic Cloud (LMC) is a disk galaxy with planar geometry and the orientation 
measurements of the LMC disk plane have been estimated previously by various authors (\citealt{df72}, 
\citealt{vc01}, \citealt{os02} and \citealt{SS10}) 
using optical data of different tracers.
Reddening plays an important role in the estimation of the structural 
parameters of a galaxy. As the effect of reddening is less in longer wavelengths, 
the LMC structure estimated using near infrared (NIR) data is likely to have reduced effect due to reddening. 
\cite{k09} (hereafter K09) derived an inclination, $\it{i}$ of 23$^o$ .5 $\pm$ 0$^o$.4 
and a postion angle of line of nodes (PA$_{lon}$), $\phi$ of 154$^o$.6 $\pm$ 1$^o$.2, using 
the JH photometric data of red clump (RC) stars 
from the Infrared Survey Facility Magellanic Cloud Point Source Catalog (IRSF MCPSC). 
This study was unable to identify the warps in the south western end 
of the disk, which is evident in the optical studies (\citealt{os02} and 
\citealt{SS10}) of the structure of the LMC using the RC stars. \\

The sample of RC stars used in the study by K09 have contamination 
from stars in the other evolutionary phases like the AGB stars. 
The details of the method applied for the reddening correction is not clearly mentioned in the 
paper by K09. The intrinsic (J$-$H)$_0$ colour of the RC stars 
estimated by K09 is in the Johnson Cousins Glass filter system. This has to be 
tranformed to the SIRIUS IRSF filter system for the accurate reddening estimation.
The extinction maps of the LMC estimated from the RC stars data in optical bands, 
given in the lower panels of Fig. 5 and Fig. 6 of \cite{SS10} and in Fig. 3 of \cite{H11} show variation in 
extinction across the galaxy. Especially a large reddening in
the south-western disk is seen in the reddening maps where warps are identified. 
Hence reddening correction is a important factor in the estimation of the structural 
parameters and the extra-planar features of the LMC.  
In the analysis done by K09, there is also an overlap of sub-regions in the peripheral regions of the disk 
which can cause some structural information to be averaged out. 
The above mentioned points motivated us to re-estimate the LMC structure using the same 
NIR data of the RC stars used by K09. In this paper we use the photometric 
data of the RC stars in the J and H pass bands from IRSF MCPSC \citep{kato07}. The K$_s$ 
band magnitude limits of the survey is too shallow to reliably detect the RC stars. 

The RC stars are core helium burning stars which are metal rich and massive counter parts 
of the horizontal branch stars. They have a constant magnitude which make 
them standard candles for distance estimation \citep{sta98}. Their constant characteristic colour 
make them good tracers for reddening estimation \citep{s05}. Here we use the JH magnitudes of RC stars 
to understand the structure of the LMC disk. The RC stars occupy a compact region in the 
optical colour magnitude diagram (CMD) which makes them easily identifiable. Even though they 
occupy a compact region in the infrared CMD, their location is blended with the location of 
stars in the other evolutionary phases. In this study, we identify the RC stars in the NIR by cross 
identifying with the optically identified RC stars to minimize the contamination of 
stars in other evolutionary phases, especially the AGB stars. 

The structure of the paper is as follows. In section 2 the data and the selection of the RC sample are 
explained. The analysis is described in section 3 and the results are given in section 4. 
The variation of structural parameters as a function of radius is described in section 5. 
The effects of reddening and population effects of RC stars in the estimation of the structure are 
discussed in section 6. 
As this work is motivated by the study of K09 and we use the same data set and the tracer that used 
in the study, in section 7 we compare our results with the study of K09 in detail. 
The differences seen are also explained in this section. The conclusions of our study are given in section 8.  
\begin{figure}
\resizebox{\hsize}{!}{\includegraphics{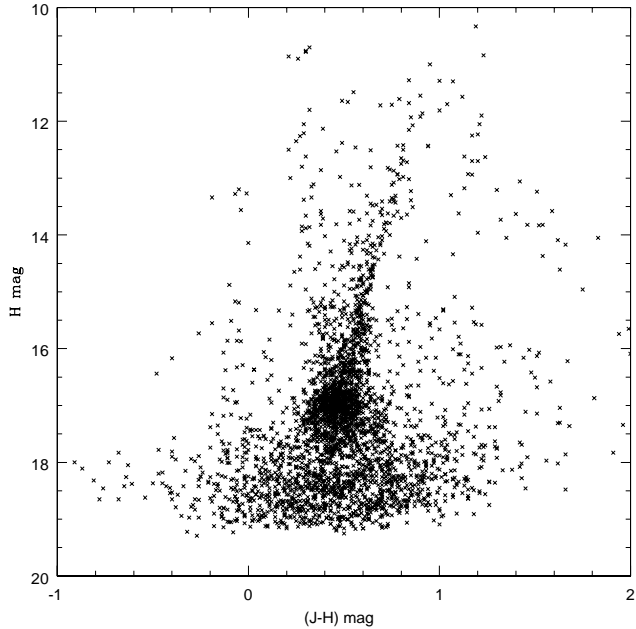}}
\caption{A sample NIR CMD of a sub-region in the LMC is shown.}   
\end{figure}
\begin{figure}
\resizebox{\hsize}{!}{\includegraphics{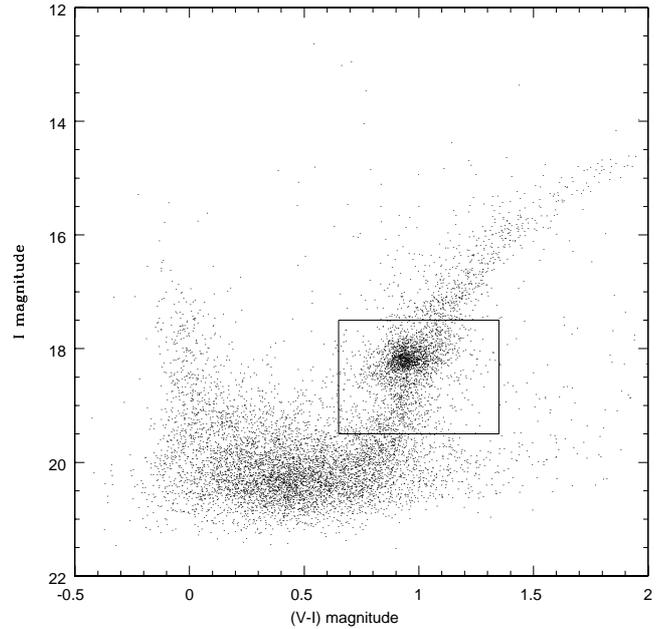}}
\caption{A sample optical CMD of a sub-region in the LMC is shown. The box used to 
identify the RC stars is also shown.}   
\end{figure}
\begin{figure}
\resizebox{\hsize}{!}{\includegraphics{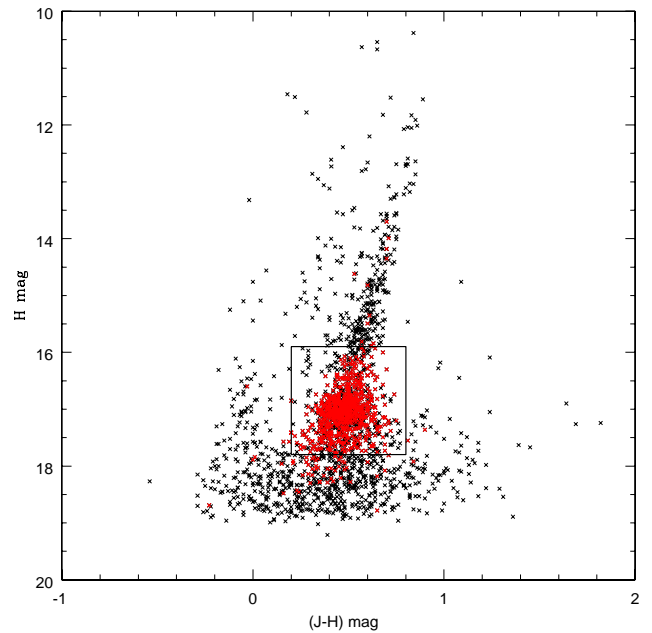}}
\caption{A sample NIR CMD of a sub-region in the LMC with the location of RC stars is shown. 
The red points are 
the RC stars identified from the optical CMD of the sub-region. The box within 
which the RC stars in the IR CMD are distributed is also shown.  
}   
\end{figure}
\section{Data} 
The IRSF Magellanic Cloud Point Source Catalog (IRSF-MCPSC) \citep{kato07} 
is an outcome of an imaging survey of the Magellanic Clouds (MCs) in 
the NIR bands J (1.25
$\mu$m), H (1.63 $\mu$m) and K$_s$ (2.14 $\mu$m) during the period October 2001 to March 2006.
The observations were made with the SIRIUS camera (Simultaneous three colour InfraRed 
Imager for Unbiased Survey) on the InfraRed Survey Facility (IRSF) 1.4 m telescope 
at Sutherland, the South African Astronomical Observatory. The SIRIUS camera is
equipped with three 1024 x 1024 HAWAII arrays to enable simultaneous observations in
the three bands (\citealt{nagashima99}, \citealt{nag03}). The IRSF/SIRIUS pixel
scale is 0.45 arcsec/pixel, yielding a field of view of 7.7 x 7.7 arcmin$^2$ . The photometric
catalog \citep{kato07} includes 14811185 point sources for a 40 deg$^2$ area of the LMC
and 2769682 sources for an 11 deg$^2$ area of the SMC. In our present study, the LMC catalog 
is used to estimate the structural parameters of the LMC using the RC stars.

We divided the IRSF MCPSC region of the LMC into 928 regions with a bin size of approximately 
10.53 x 15 arcmin$^2$ . The average photometric error (magnitude range 15 - 20 in 
J and H bands) is around 0.1 mag. Photometric data with error less than 0.3 mag are considered 
for the analysis. A sample $(J-H)$ vs H CMD 
is shown in Fig. 1. To isolate the approximate RC location in the infrared CMD, we
used the optical CMD of the corresponding sub-region. The RC stars are easily 
identifiable in the optical CMD as a separate component.\\

The Magellanic Cloud Photometric Survey (MCPS) (Zaritsky et al. 1997) obtained
the UBVI photometry of virtually all stars brighter than V= 21 mag in the MCs. 
The five year survey was conducted at the Las Campanas Observatory$'$s 1 m Swope telescope and
the images were obtained using the Great Circle Camera (GCC, Zaritsky et al. 1996). The
thinned 2048 x 2048 CCD has 0.7 arcsec/pixel scale. The survey scanned 64 deg$^2$ of the
LMC and 16 deg$^2$ of the SMC. The IRSF observed region of the LMC comes within the MCPS observed 
region of the LMC. Zaritsky et al. (2004) presented the
data of the LMC MCPS survey. The MCPS observed region of the LMC is also sub-divided into 
1512 regions (each with an area of 10.53 x 15 arcmin$^2$). For each sub-region (V$-$I) vs I 
CMD is plotted and the RC stars are identified. A sample optical CMD of the LMC is shown
in Fig. 2. For all the regions, the RC stars are found to be located well within the box shown 
in the CMD, with boundaries 0.65 - 1.35 mag in (V$-$I) colour and 17.5 - 19.5 mag in I magnitude.
The average photomeric error of stars in the RC magnitude range is 0.05 mag in the V and I bands. 
We considered only stars with error less than 0.15 mag for the identification of the RC stars.\\

We cross-identified the optical and infrared data and obtained the infrared (J,H) 
magnitudes of the RC stars identified within the box of the optical CMD 
(Fig. 2). Those are shown as red points in Fig. 3. 
Most of the RC stars are well within the box of infrared CMD, 
with boundaries 0.2 - 0.8 mag in $(J-H)$ colour and 15.9 - 17.8 mag in H magnitude.  
Stars outside the box are not considered for the analysis. There are a few 
stars outside the box, especially fainter than 17.8 mag. As the H band limiting magnitude 
of the survey is 17.8 mag, it is justifiable to exclude those stars from the analysis. 
Again, from Fig. 3 we can see that the peak of the RC stars in H band 
is brighter than 17.8. As our method needs to identify the peak of the RC distribution, the box is 
used. The RC stars within the box are optically identified for all the 928 sub-regions. 
Out of 928 regions only 926 regions have reasonable number (100-3000) of RC stars to do the analysis. 

\section{Analysis}

\begin{figure}
\resizebox{\hsize}{!}{\includegraphics{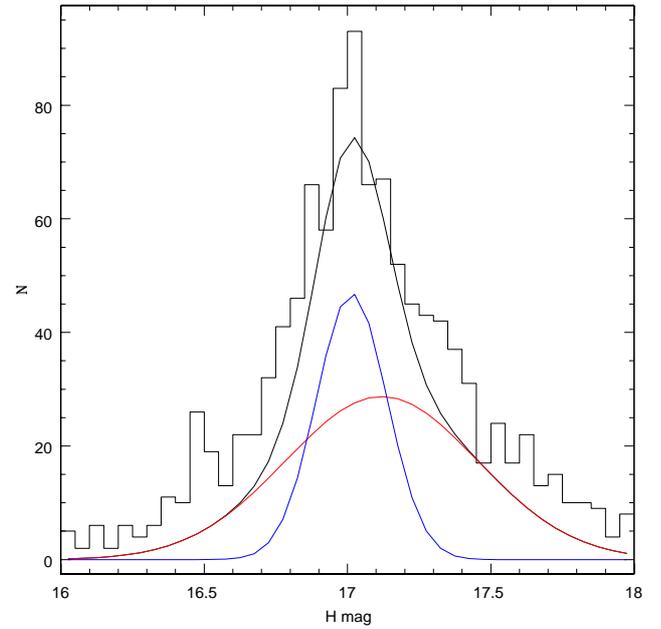}}
\caption{A typical H magnitude distribution of a sub-region in the LMC. 
Each Gaussian profiles are shown as red and blue lines. The combined best fit profile 
of double Gaussian is shown as black line. The reduced $\chi^2$ value of the fit is 1.1. 
}   
\end{figure}

\begin{figure}
\resizebox{\hsize}{!}{\includegraphics{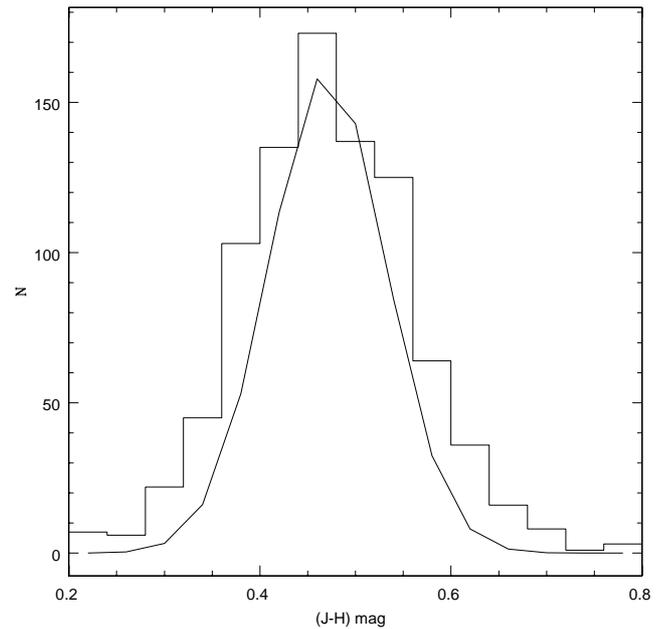}}
\caption{A typical $(J-H)$ colour distribution of a sub-region in the LMC. The best fit 
profile to the distribution is also shown. The reduced $\chi^2$ value of the fit is 1.2. 
}   
\end{figure}

\subsection{Identification of the peak magnitude and colour of the RC distribution}

\subsubsection{Method 1}
The number distribution of the RC stars 
in H magnitude and $(J-H)$ colour are obtained with a bin size of 0.05 mag and 0.04 mag 
respectively. The obtained distributions in colour and magnitude are
fitted with a) a Gaussian function, b) Gaussian + quadratic polynomial and c) Combination of two 
Gaussian functions. A non linear least square method is used for fitting and the parameters 
are obtained. The parameters obtained are the coefficients of each term 
in the function used to fit the profile, error in the estimation of each parameter and
reduced $\chi$$^2$ value. The errors in the estimated parameters are calculated using the
covariance matrix. By comparing the reduced $\chi^2$ values of different profile fits we found that combination 
of two Gaussian profiles, one narrow component and one broad component, fits well for the H magnitude
distribution.  
The $(J-H)$ colour distribution in majority of the ($\sim$ 60 $\%$) sub-regions 
fits better with a single Gaussian. The colour distributions in the remaining sub-regions are best fitted   
with a double Gaussian distribution or a Gaussian + quadratic polynomial. By comparing 
the reduced $\chi^2$ values, the best profile fit is used for further analysis.  
The H magnitude and $(J-H)$ colour distributions are shown in Fig. 4 and Fig. 5 respectively. 
The best fit profiles are also shown in the figures. From Fig. 4 we can see that the H 
magnitude distribution is well represented by two Gaussian functions. Each Gaussian profile is separately shown as 
blue and red lines and the combined profile is shown as black line. 
The peak of the narrow component coincides with the the bin with maximum number of stars. Thus 
this value is taken as the RC peak magnitude for our analysis. The peak shift between the broad and narrow component 
ranges from 0.05 mag to 0.1 mag. The broad component peak is fainter than the narrow component peak in some 
sub-regions and in some other sub-regions it is brighter. The broad component may be representing the thick disk 
RC population and/or can be due to stars in other evolutionary phases which are the contaminants in the sample. 
The NIR sample are optically selected and in optical magnitude distribution such a double peak feature is 
not seen \citep{SS10}. Hence we cannot conclusively say anything about the broad component peak and dispersion. 
In Fig. 5, the RC colour distribution which is best fitted by a single Gaussian is shown. In 20$\%$ of the 
sub-regions, the RC colour distribution shows double peak with broad and narrow components. 
The peak shift between the broad and narrow components is not always in one direction. The broad component 
peak is bluer than the narrow component peak in some sub-regions and 
in some other sub-regions it is redder. The narrow component peak is taken as the RC peak colour. 
Note that the RC colour distribution in majority of sub-regions has only one peak.
The parameters that are needed to estimate the 
structure of the LMC are the peak H mag, peak $(J-H)$ mag, the associated errors 
and the reduced $\chi$$^2$ values. Regions with peak errors greater than 0.1 mag and those with 
reduced $\chi$$^2$ value greater than 3.0 are omitted from the analysis. 
Thus the regions used for final analysis reduced to 775 from 926. Around 85 $\%$ 
of the sub-regions are available for the final analysis. 
\subsubsection{Method 2}
In the method 1, 15 $\%$ of the sub-regions of the LMC are omitted from 
the final analysis due to poor fit of the RC distribution. Many sub-regions in the central region 
(within in a radius of 2.5 degrees) are part of the omitted data points. Some of the outer regions are 
also omitted. The structural parameters of the 
LMC are very much dependent on the coverage and hence these inner regions which are omitted can make a 
vast difference in the estimates \citep{SS10}. The location of the bar with respect to the disk and 
the structural parameters of the inner region (within the radius of $\sim$ 2.5-3 degrees) are some of 
the interesting aspects related to the structure of the LMC.  Thus to understand the structure of the LMC 
in more detail, it is important to increase the available regions in the inner region. Though the 
most appropriate method for the estimation of the accurate peak of the RC distribution in a sub-region 
is by fitting a profile by numerical analysis (method 1), we can also estimate the peak of the 
distribution as the average of the magnitudes of stars in the bin with largest number of stars. This 
allows us to estimate the peak mag of RC stars in all the sub-regions including the central regions. 

The RC stars identified are binned in both H mag and (J$-$H) colour with a bin size of 0.04 and 0.03 mag 
respectively. The bin with largest number of RC stars is identified in both the magnitude and colour distribution. 
The average H mag, (J$-$H) colour of the stars in that particular bin and the associated standard deviation 
are estimated. In order to reduce the effect of binning in the estimation of average H mag and (J$-$H) colour, 
we identified the number of RC stars in each bin on either sides of the bin with largest number of RC stars. 
If the number of RC stars in those bins are greater than the number, (N $-$ $\sqrt N$) where N is the largest 
number of RC stars identified, those stars are also included in the estimation of average values.  

\subsection{Estimation of E(J$-$H) reddening}
To estimate the structural parameters of the LMC disk from the RC magnitude we need to correct for 
extinction. The RC peak $(J-H)$ mag at each location is used to estimate the E(J$-$H) reddening and hence the 
extinction in H band. The reddening is calculated using the relation, \\

E$(J-H)$ = $(J-H)_{obs}$ $-$ $(J-H)_{intrinsic}$\\

To obtain the absolute J and H band RC magnitudes and hence the $(J-H)_{intrinsic}$ colour of RCstars in the LMC, 
we used the method 1 described in \cite{GS01}. The mean RC properties as a function of time and metallicity 
based on \cite{g00} isochrones are available in table format in http://pleiadi.pd.astro.it. We used 
the star formation rate from \cite{hz09} and the age metallicity relation given by \cite{PT98} to 
obtain the absolute mean absolute J and H magnitudes of RC stars in the LMC. The RC stars in the LMC has an age 
range. They are older than 1 Gyr and younger than 10 Gyr. We used 1-10 Gyr age range and also 1.5-9.5 Gyr age 
range and obtained the mean magnitudes. The mean and the standard deviation of the values obtained in the above 
mentioned two age ranges are used as the final value. The (J$-$H)$_{intrinsic}$ value is obtained as 
0.47 $\pm$ 0.02 
mag. K09 did a similar analysis to estimate the intrinsic colour of RC stars, assuming a constant star 
formation rate and a slightly different age range. The value obtained by K09 for (J$-$H)$_{intrinsic}$ is also 
0.47 $\pm$ 0.06 mag which matches with our estimate. 

But the Girardi isochrones estimate the absolute magnitudes in \cite{bb88} filters. The IRSF SIRIUS 
observations are done in MKO filter system. Thus it is important to do the necessary transformations to obtain 
the (J$-$H)$_{intrinsic}$ value in MKO system. The (J$-$H)$_{intrinsic}$ value in the MKO system has to be used for the 
reddening correction. We did the necessary transformations \citep{car01} and obtained the  
(J$-$H)$_{intrinsic}$ colour in MKO system as 0.40 $\pm$ 0.03 mag. The errors in the transformation co-efficients are also 
considered while estimating the final error associated with the (J$-$H)$_{intrinsic}$ colour in MKO system. In the process, 
the absolute mean H and J magnitudes of RC stars in the LMC are also estimated. The mean values are M$_J$ = -1.13 
$\pm$ 0.02 mag and M$_H$ = -1.53 $\pm$ 0.02 mag. 

The reddening, \\

E(J$-$H) = $J-H_{observed peak}$ $-$ 0.40 $\pm$ 0.03 mag\\

towards each sub-region is estimated. The interstellar extinction towards each sub-region 
is estimated using the relation,\\

A$_H$ = 1.65 $\pm$ 0.16 x E$(J-H)$ (K09).
\subsection{Estimation of the relative distance of sub-regions from the center}
The dereddened H$_0$ magnitude of the RC stars for each sub-region is estimated using H$_0$ = H $-$ A$_H$. 

The difference in H$_0$ mag between regions is assumed only due to their difference in the relative distance, 
$\Delta$D in kpc, calculated using the distance modulus formula given below.\\ 

$\Delta$D = 50 x 10$^{(H_0{region} - H_0{mean})/5}$, \\ 

where 50 is the distance to the center of the LMC in 
kpc.\\

\subsection{Plane fitting procedure}
The R.A, Dec and relative distance of a sub-region from the center are used to create a 
cartesian coordinate system using the transformation equations given below (\citealt{vc01}, 
\citealt{SS10}  
see also Appendix A of \citealt{wn01}).\\\\
  x = -Dsin($\alpha$ - $\alpha_0$)cos$\delta$,\\\\
  y = Dsin$\delta$cos$\delta_0$ - Dsin$\delta_0$cos($\alpha$ - $\alpha_0$)cos$\delta$,\\\\
  z = D$_0$ - Dsin$\delta$sin$\delta_0$ - Dcos$\delta_0$cos($\alpha$ - $\alpha_0$)cos$\delta$,\\

 where D$_0$ is the distance to the center of the LMC and D, the distance to the each sub-region 
is given by D = D$_0$ + $\Delta$D.
The ($\alpha$, $\delta$) and ($\alpha_0$ , $\delta_0$) represents the R.A and Dec of the 
  region and the center of the LMC respectively. In our analysis, the optical center of the LMC, 
$05^h19^m38^s.0$ $-69^o27'5".2$ (J2000) (\cite{df72}) is taken as the center of the LMC. According to 
\cite{vc01} the adopted center does not affect the derived parameters for the LMC disk plane 
(inclination, PA$_{lon}$).The distance, 
D$_0$ to the center of the LMC is taken as 50 kpc. 

Once we have the x,y, and  z coordinates we can apply a weighted least square
plane fit to obtain the structural parameters of the LMC disk. 
The equation of the plane used for the plane fit is given by \\

     z = Ax + By + C\\

From the coefficients of the plane A,B \& C,  $\it{i}$ and  $\phi$ can be calculated using the 
formula given below.\\

    Inclination, ${i}$ = arccos(C/$\sqrt{A^2+B^2+1}$) \\
    
    PA$_{lon}$, $\phi$ = arctan(-A/B)+sign(B)$\pi$/2.\\

We calculated the deviations of the LMC disk 
from the plane with estimated coefficients. 
The expected z for a plane is calculated from the 
equation of a plane. The difference in the expected and 
calculated z values is taken as the deviation of the LMC disk from the 
plane. Thus the extra-planar features of the LMC disk are identified 
and quantified. Once the deviations are estimated, the regions with 
deviations above the error in z are omitted and the plane-fitting 
procedure is applied to the remaining regions to re-estimate the 
structural parameters of the LMC disk plane.\\

The uncertainty in the H and $(J-H)$ peak values, uncertainty in the intrinsic colour of the RC stars, 
uncertainty in the reddening law are all propagated properly to estimate the error in relative distance. 
The error in the z values are used in the weighted least square plane fitting procedure to obtain the error 
in the estimated disk parameters. Thus the error in the estimate of the LMC disk parameters is calculated 
by propogating the systematic errors associated with all the quantities involved in the estimation.

\begin{figure}
\includegraphics[width=10cm, height=8cm]{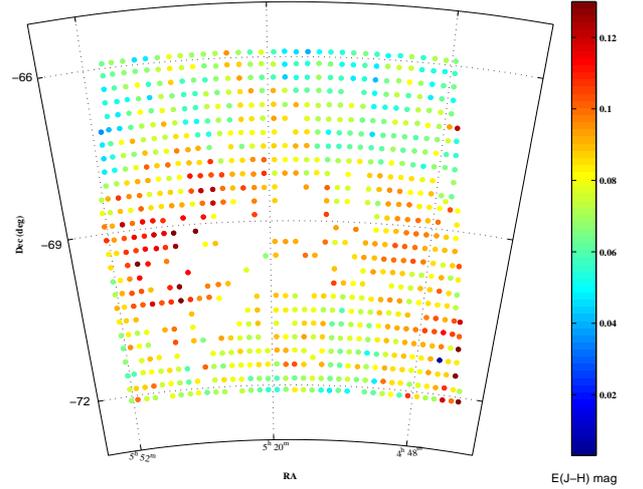}
\caption{Two dimensional plot of reddening (E(J$-$H)) obtained based on method 1.}
\end{figure}

\begin{figure}
\includegraphics[width=11cm, height=7cm]{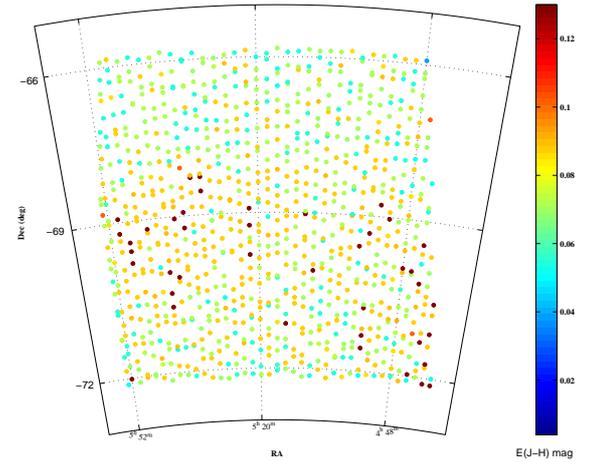}
\caption{Two dimensional plot of reddening (E(J$-$H)) obtained based on method 2.}
\end{figure}

\section{Results}

\subsection{NIR Reddening Map towards the LMC}
One of the by products of this study is the NIR reddening map towards the LMC. The shift in the peak of the 
(J$-$H) colour distribution with respect to the characteristic (J$-$H) colour of RC stars is a measure of 
reddening. The peak (J$-$H) colour of the RC stars in each sub-region is estimated based on method 1 and 
also based on method 2. The intrinsic value of the E(J$-$H) colour of the RC stars in the LMC is chosen as 
0.40 $\pm$ 0.03 mag. Using this value, the E(J$-$H) value of each sub-region is estimated as described in 
section 3.2. A colour coded two dimensional reddening map obtained based on method 1 is shown in Fig 6. 
The E(J$-$H) value has a range from 0.04 mag to 0.13 mag with an average of 0.08$\pm$0.03 mag. 
A similar plot obtained based on method 2 is 
shown in Fig 7. The E(J$-$H) value has a range from 0.05 mag to 0.15 mag with an average of 0.08$\pm$0.03 mag. 
From the figures, Fig 6 and Fig 7 we can see that there are slight variations in the reddening estimates 
when region to region is considered. But the locations where large/less reddening are seen coincide well. 
Also, the range of reddening values (0.04-0.13 mag and 0.05-0.15 mag) and the average reddening (0.08 $\pm$ 0.03 mag) estimated from 
both the methods match well. In order to get a quantitative estimate of the variations, the difference in the 
reddening estimates obtained for each region using two methods is calculated. The difference has a range from 
a minimum value of 0.0002 mag to a maximum value of 0.05 mag. In most of the regions, 
the difference is $\le$ 0.009 mag , which is much less than the average error (0.03 mag) associated with the 
estimation of reddening in both the methods. Only for a few regions (less than 6\%) the difference is 
more than 0.03. From these maps we can see that reddening varies across the LMC. 
The reddening is more in the south western regions of the LMC disk as well on the eastern side. 
The reddening variations obtained here are similar to those obtained in the optical studies 
(\citealt{H11} \& \citealt{SS10}).

\begin{figure*}
\resizebox{\hsize}{!}{\includegraphics{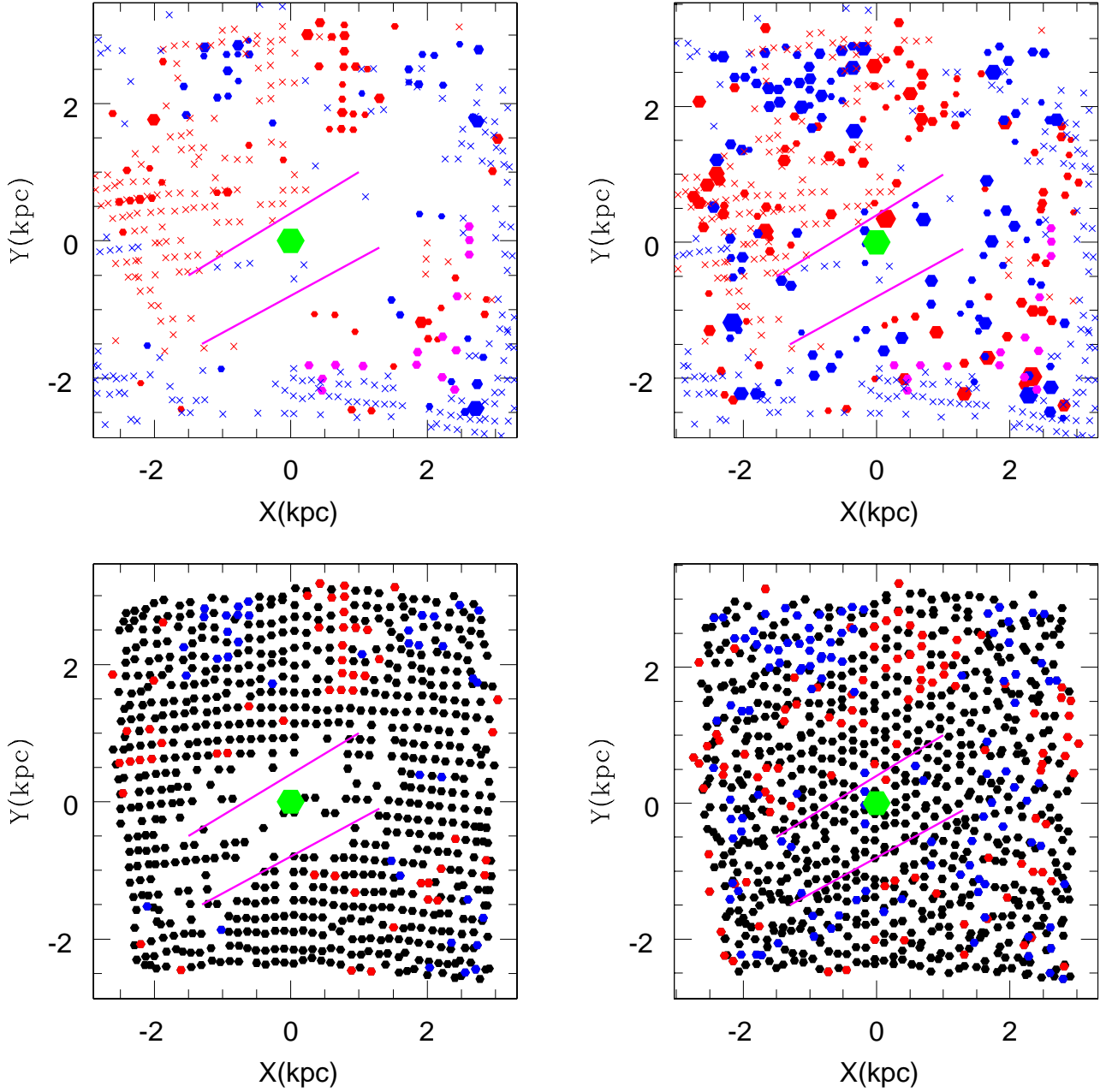}}
\caption{The IRSF MCPSC regions which are fitted on the plane and those which 
are deviated are shown. The lower left and upper left panels are the plots which are obtained from the 
analysis using method 1. The lower right and upper right panels are the plots obtained from the 
analysis using method 2. In the lower panels, black dots represent regions on the fitted LMC
plane, red dots represent regions behind the fitted plane and 
the blue dots represent regions which are in front of the fitted 
plane. The upper panels show only regions with deviations, greater than 
1.5 kpc out of the plane and the size of the points are proportional 
to the amplitude of the deviations. The blue and red crosses are 
regions which are infront of the plane and behind the plane  
respectively, identified in the optical study using MCPS data \citep{SS10}. 
Magenta dots are regions which are suggested as warps by \cite{os02}.
The green hexagon in both the panels represents the optical center of the 
LMC.}
\end{figure*}

\subsection{Structural parameters of the LMC disk}

\subsubsection{Method 1}

The structural parameters of the LMC disk are estimated using the dereddened mean 
H$_0$ magnitude
of the RC stars in 775 regions. This method gives an inclination of i= 26$^o$.0 $\pm$ 1$^o$.2 \& PA$_{lon}$ 
= 146$^o$.5$\pm$3$^o$.5 for
the LMC disk. The deviation of the LMC disk regions from the estimated plane are calculated as
explained in section 3.3. 
Deviations above 1.5 kpc are considered as significant deviations
from the fitted plane. The lower left and upper left panels of Fig 8 show the deviation of the LMC regions from 
the plane. In the lower left panel, all the regions used for the analysis are plotted. The black points are those which are on the
fitted plane, red points are disk regions which are behind the plane and blue points are the disk
regions which are in front of the plane. In the upper left panel only the regions with deviations above
1.5 kpc are plotted and the size of the points are proportional to the amplitude of the deviation.
Here also, red points are regions behind the fitted plane and blue are in front of the fitted plane. 
The magenta points are the regions where warps are identified by \cite{os02}. The blue and red  
crosses are the regions where infront of the plane and behind the plane features are identified in 
the study of \cite{SS10} using MCPS data. From the plots we can see that there are some regions 
which are deviated from the planar structure of the LMC disk. Even though the coverage of MCPS is 
larger than that of the IRSF MCs survey and that there are large number of regions omitted in the IR analysis, 
we can see that most of the locations of deviations identified in the present study match with the locations of 
deviations identified from the previous optical analysis \citep{SS10}. 
In the south western LMC, there are regions which are closer to us and also away from us.

The presence of extra-planar features would affect our estimation of the structural 
parameters of the disk. Out of 775 regions, 91 regions show deviations greater than 1.5 kpc  
and are considered as real deviations. We removed these regions and re-estimated the structural 
parameters of the disk using the remaining 684 regions. We obtained an inclination of i= 26$^o$.6 
$\pm$ 1$^o$.3 \& PA$_{lon}$ = 148$^o$.3$\pm$3$^o$.8. The dereddened RC magnitude is plotted against 
the axis perpendicular to the line of nodes which is the axis of maximum gradient and it is shown in the lower 
panel of Fig 9. 
The plot clearly shows the effect of inclination from NE to SW of the LMC disk. The black points are those on the 
plane of the disk and the red are the extra-planar regions. The slope and y-intercept of the line fitted to 
the data, excluding the extra-planar regions are 0.019 $\pm$ 0.001 mag/kpc and 16.94 $\pm$ 0.1 mag. The slope is the measure 
of the inclination and the inclination estimated from the slope is 25$^o$.4$\pm$1$^o$.4.

\subsubsection{Method 2}
Using this method we obtained an inclination of i= 26$^o$.6 $\pm$ 1$^o$.4 \& PA$_{lon}$ = 
138$^o$.7$\pm$3$^o$.5 
for the LMC disk from the analysis of 919 sub-regions. The standard deviations estimated for the 
peak H mag and $(J-H)$ colour are used to estimate the error in the estimation of distance. 
The regions which show deviation larger than 1.5 kpc are considered as real deviation. 
 The lower right and upper right panels of Fig 8 show the deviation of the LMC regions from 
the plane. The optically identified deviations are also plotted. The symbols are the same as 
shown in the lower left and upper left panels of Fig. 8. Most of the deviations identified are similar to that 
obtained based on method 1. As there are more regions in the analysis based on method 2, there are some more 
extra-planar features identified. These deviations more or less match with the optically identified 
deviations. The structures in the south western region are better revealed in this analysis. As seen earlier, 
in the south western region along with the regions which are closer to us (which matches with the regions where 
warps are identified by \citealt{os02}), there are regions which are away from us as well.  Based on the 
amplitude of the deviations in the south western region, the deviations closer to us are significant than those  
away from us. Out of 919 regions, 215 regions show deviations. These regions are removed and the 
structural parameters are re-estimated. An inclination of i = 25$^o$.7$\pm$1$^o$.6 and 
PA$_{lon}$ = 141$^o$.5$\pm$4$^o$.1 are obtained. The inclination obtained matches well 
within errors with the inclination obtained based on method 1. The PA$_{lon}$ is slightly different. As the 
estimated parameters very much depend on the coverage of the data set, the variation  
in these values can be attributed to more number of inner and outer regions 
included in the analysis based on method 2 as compared to method 1. The dereddened H 
magnitude is plotted against 
the axis of maximum gradient and is shown in the upper panel of Fig. 9. The colour scheme is the same as that 
of the plot in the lower panel of Fig. 9. From 
the slope and intercept of the line fitted to the data points, the inclination of the LMC disk and the distance 
modulus to the LMC center are estimated. We estimated an inclination of 25$^o$.0$\pm$1$^o$.5 for the LMC disk.  

The significant result of this analysis, based on method 2, is that the structure of the central region  
is revealed. Most of the regions in the bar are on the plane and there are only a few regions which show 
deviations from the plane. This suggests that the bar is co-planar with the disk. This result is very similar to 
the result obtained by \cite{ss09apj} regarding the location of the bar with respect 
to the disk. Even though most of the regions in the bar are on the plane of the LMC disk, there are some 
localized regions which are brighter than the nearby regions. There is also one region near the center which is 
away from us. These suggest that the bar region of the LMC is structured. 

\begin{figure}
\resizebox{\hsize}{!}{\includegraphics{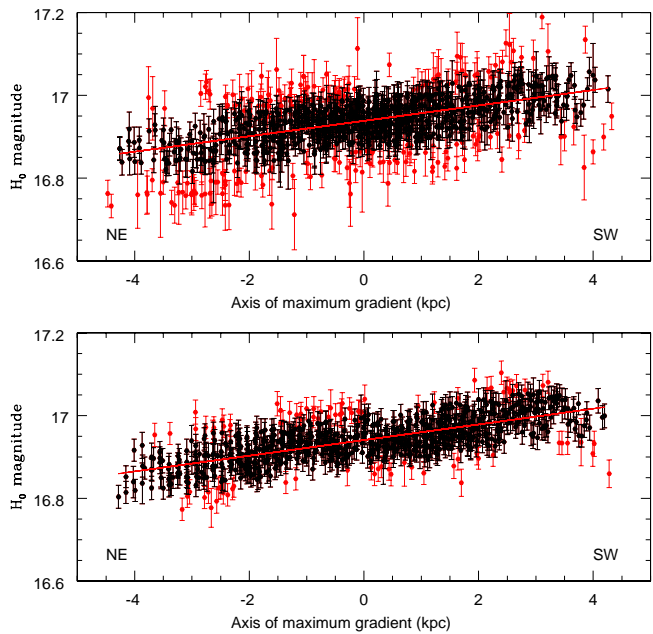}}
\caption{Dereddened RC magnitude plotted against the axis of maximum   
gradient.  The RC magnitude estimated from method 1 and method 2 are plotted in lower and upper panels 
respectively. The red points are regions which show deviation larger sigma.
 The direction of inclination is shown as red line.}
\end{figure}

\subsection{Distance Modulus}
The y-intercepts obtained from the Fig 9 and Fig 11 are the mean H$_0$ value of the RC stars in the LMC. 
The mean distance modulus to the LMC can be obtained using the formula,\\ 

$\mu_0$ = $H_0$$_{mean}$ - $M_H$$_{(LMC)}$. \\ 

In the above equation, M$_H$ is the absolute H band magnitude of the RC stars 
in the LMC. M$_H$ is taken as -1.53 $\pm$ 0.02 which is calculated in section 3.2. 
The H$_0$$_{mean}$ obtained from the lower panel of Fig 9 (method 1) 
is 16.94$\pm$0.1 mag and from the upper panel of Fig 9 (method 2) is 16.95$\pm$0.1 mag. Then the mean distance 
modulus 
to the LMC, $\mu_0$ is 18.47$\pm$0.1 mag (from method 1) and 18.48$\pm$0.1 mag (from method 2). These values match 
with in errors with the previous estimates of 18.54$\pm$0.06 mag (K09), 18.55$\pm$0.01 mag \& 
18.5$\pm$0.01 mag (from MCPS and OGLE III data sets, \citealt{SS10}), 
18.5 $\pm$ 0.02 (\citealt{A04}) and 18.53 $\pm$ 0.07 (\citealt{SG02}) towards the LMC.

\section{Dependence of structural parameters on the photometric band}
In the present study to estimate the structural parameters of the LMC disk, we assume that the variation in the 
extinction corrected H band magnitude of the RC stars between various sub-regions is solely due to the distance 
effect induced by the structure of disk. Then the variations should be independent of the chosen photometric bands. 
Thus, it is useful to compare the results obtained from the magnitude distribution of RC stars in H band with 
that of J band analysis. The analysis in J band described below is done similar to method 2 given in section 3.1.2 
for H band analysis. The optically identified RC stars are over plotted in the J vs $(J-H)$ CMD and is shown in 
Fig 10. From the 
CMD we can see that most of the RC stars are well within the box of infrared CMD, with boundaries 0.2 - 0.8 mag in 
$(J-H)$ colour and 16.4 - 18.4 mag in J magnitude. For all the 926 sub-regions, the RC stars are identified with 
in this box. The RC stars identified are binned in J magnitude with a bin size of 0.04 mag. 
The bin with largest number of RC stars is identified in J magnitude distribution. 
The average J mag in that particular bin and the associated standard deviation 
are estimated. In order to reduce the effect of binning in the estimation of average J mag,  
we identified the number of RC stars in each bin on either sides of the bin with largest number of RC stars. 
If the number of RC stars in those bins are greater than the number, (N $-$ $\sqrt N$) where N is the largest 
number of RC stars identified, those stars are also included in the estimation of average values. The E(J$-$H) 
value estimated for each sub-region in section 4.1 is used to find the extinction in J band for that region. 
The formula used is \\

A$_J$ = 2.63 $\pm$ 0.23 * E(J-H) \citep{in05}\\

The extinction corrected J band magnitude for all the sub-regions are estimated. From the extinction corrected J 
band magnitudes the relative distance between regions and hence the z co-ordinates are obtained. A weighted least 
square plane fitting procedure is applied to estimate the structural parameters of the LMC disk. The deviations 
from the plane are also estimated. The structural parameters are re-estimated after removing the regions which show 
deviations larger than 1.5 kpc from the plane. An inclination of i = 24$^o$.4$\pm$2$^o$.5 and 
PA$_{lon}$ = 146$^o$.3$\pm$6$^o$.5 are obtained. The parameters match well with in errors with the estimates from 
the analysis of H band data. The deviations from the plane are shown in Fig 11. The black points are those on 
the plane. The red and blue are points are regions which are behind and in front of the plane respectively. The size 
of the red and black points is proportional to the amplitude of deviations. The deviations obtained also match 
well with the deviations (shown in Fig 8) obtained from the analysis of H band data. 
The distance modulus towards the LMC is also calculated using the J band data and is found to be 
18.47 $\pm$ 0.1 mag. The results for the analysis of J band data match well with the results obtained from the 
analysis of H band data.

\begin{figure}
\resizebox{\hsize}{!}{\includegraphics{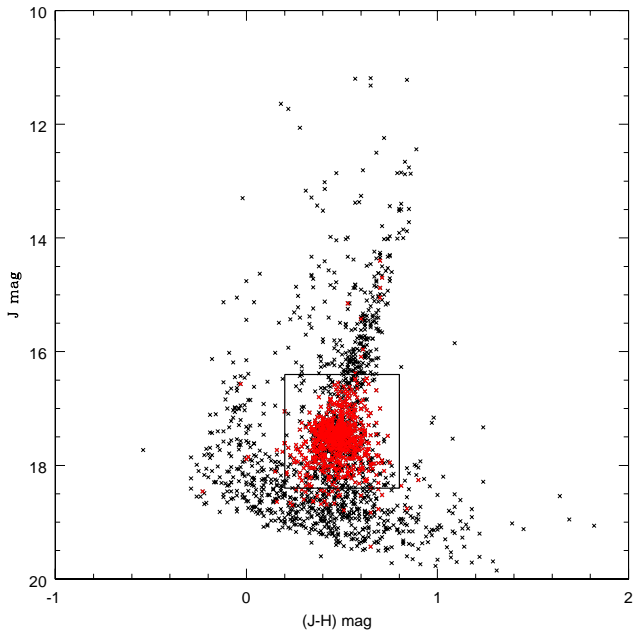}}
\caption{A sample $(J-H)$ vs J CMD of a sub-region in the LMC with the location of RC stars is shown. 
The red points are 
the RC stars identified from the optical CMD of the sub-region. The box within 
which the RC stars in the IR CMD are distributed is also shown.  
}   
\end{figure}

\begin{figure}
\resizebox{\hsize}{!}{\includegraphics{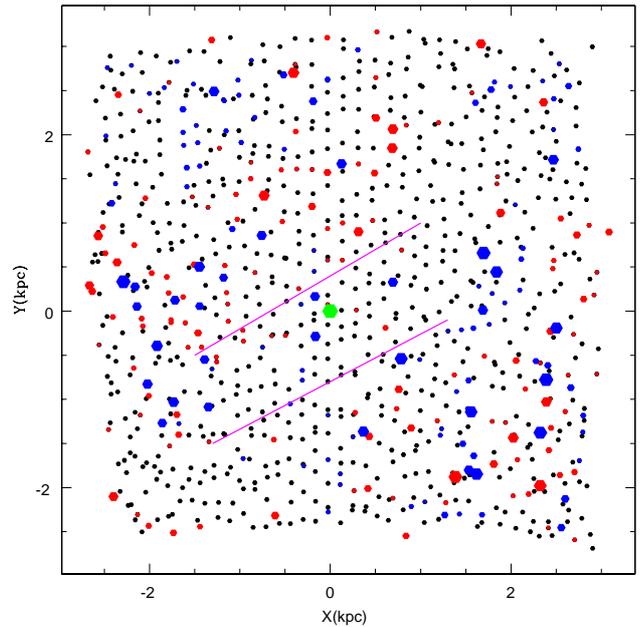}}
\caption{The IRSF MCPSC regions which are fitted on the plane and those which 
are deviated based on the analysis of J band data are shown. The black dots represent regions on the fitted LMC
plane, red dots represent regions behind the fitted plane and 
the blue dots represent regions which are in front of the fitted 
plane. The size of the blue and red points are proportional to the amplitude of the deviations.}
\end{figure}

\begin{figure}
\resizebox{\hsize}{!}{\includegraphics{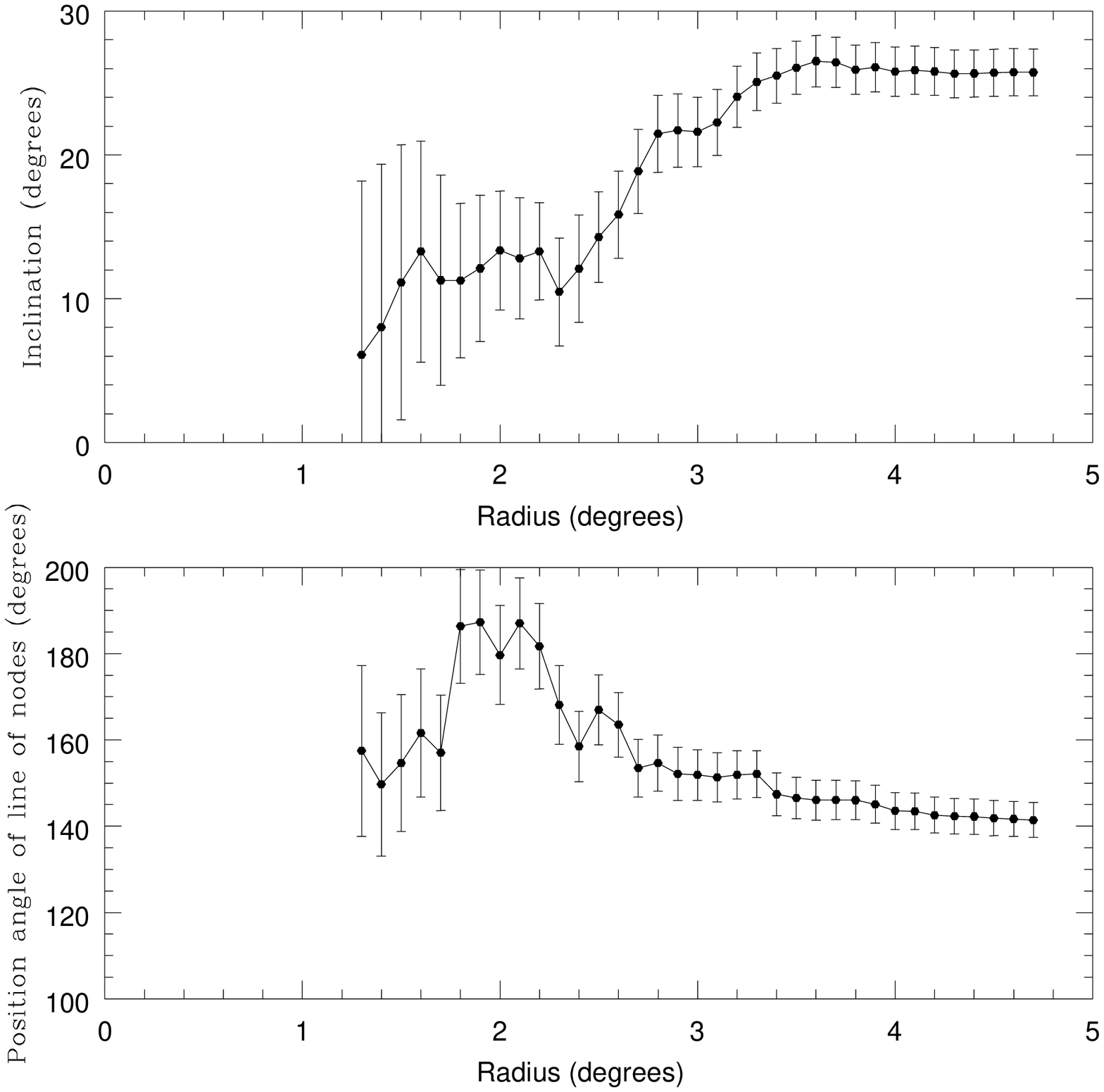}}
\caption{The variation of the structural parameters of the LMC disk as a function of radius is shown. The 
upper panel shows the variation of inclination and the lower panel shows the variation of the PA$_{lon}$.}
\end{figure}

\section{Radial Variation of the structural parameters} 


In this section, the variation of the structural parameters of the LMC disk as a function of radius is studied. 
As the data points are more in the analysis based on method 2, we make use of that data to find the 
variation of the parameters as a function of the radius. We estimated the parameters using 
the data within different radii, starting from 1.5 degrees to 4.5 degrees with an increment of 0.1 degree. 
The estimates obtained are 
plotted in Fig. 12 as function of radius. The lower panel shows the variation of PA$_{lon}$ and the upper panel 
shows the variation of inclination. The parameters are more or less similar above the radius of 3 degrees. 
From the plot we can also see that the inner LMC, (within the radius of 3 degrees) 
is disturbed. Inclination of the disk increases gradually from inner to outer region till a radius of $\sim$ 
3.25 degrees. The PA$_{lon}$ varies much within a radius of 2.8 degrees 
and then remains almost constant outwards. 
Variations in the inclination and position angle of line of nodes are seen with in $\sim$ 3 degree radius. 
Variations with in the inner region suggest that the  inner LMC, where the bar is also located, 
is structured/disturbed. This could be due to the the effect of tidal interactions and/or mergers experienced 
by the LMC. 
After 3.25 degrees, both the parameters of 
the disk remain almost constant ($\it{i}$ $\sim$ 26$^o$.0 and PA$_{lon}$ $\sim$ 143$^o$.0). 
\cite{SS10} estimated the parameters of the outer 
disk (regions which have radius greater than 3$^o$) and that of the inner disk 
(data within in the radius of 3$^o$). They found the inner disk to have lower inclination 
and larger PA$_{lon}$ and vice versa for the outer disk. As they had only a few regions in the inner 3 degree 
radius they could not see the continuous trend from inner to outer region. Here from this analysis, we confirm 
that the inner structure of the LMC is different from the outer structure and the inner disk is less 
inclined with large PA$_{lon}$ and the outer disk is highly inclined with less PA$_{lon}$. 
The increase in the inclination of the outer disk, which makes the north eastern part of the LMC more closer 
to us, could be due to the tidal interaction with the our Galaxy.
\begin{figure}
\resizebox{\hsize}{!}{\includegraphics{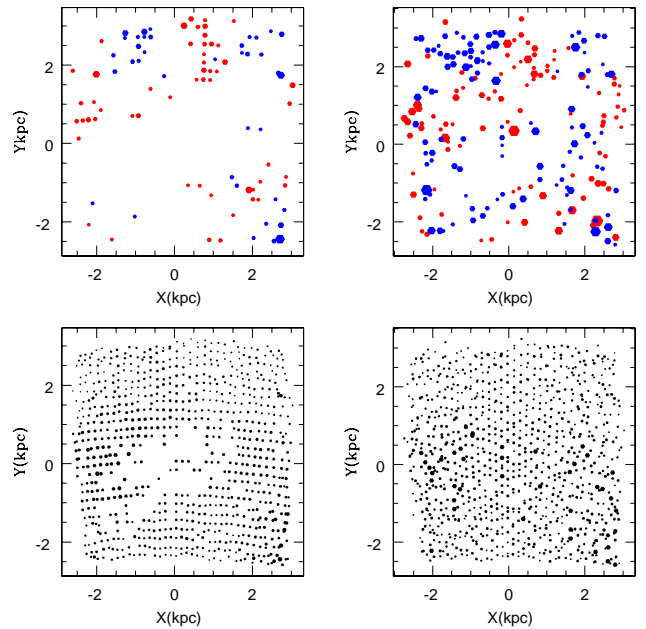}}
\caption{The lower left and lower right panels show the distribution of reddening across the LMC, estimated  
using method 1 and method 2 respectively. The size of 
the points is proportional to the amplitude of the reddening. The upper left and upper right panels show the 
distribution of deviations from the plane estimated using method 1 and method 2 respectively. 
The blue points are the regions which are 
infront of the plane and the red points are those which are behind the plane.}   
\end{figure}

\section{Effect of reddening and population effects in the detection of extra-planar features}
The extra-planar features identified in the disk of the LMC are important as they 
give clues to the interaction of the LMC with the external systems. The effects of the 
population differences of the RC stars and reddening in the detection 
of the extra-planar features have to be looked at carefully to understand whether the 
structures identified are real or not. In this section we discuss in detail the 
above mentioned effects. 

\subsection{Heterogeneous population}
The RC stars in the LMC disk are a heterogeneous population, and therefore  
they would have a range in mass, age, and metallicity. The density of stars in 
various locations will also vary with the local star-formation rate as a 
function of time. These factors result in a range of magnitude and color of 
the net population of RC stars in any given location and would contribute to 
the observed peaks in magnitude and color distributions. Therefore, the deviations 
found in some regions may also be due to these population differences of RC stars. 
But previous studies by \cite{os02}, \cite{vc01}, \cite{sa02}, \cite{G06} and \cite{pia13} suggest 
that the population effects in the magnitude and colour of the RC stars in the 
central regions of the LMC are likely to be negligible. \\

The contamination of the RC sample by stars in the other evolutionary phases, like the AGB stars, 
is less in our analysis as the stars in our sample are optically identified RC stars, as described 
in Section 2.  

\subsection{Effect of Reddening}
The extra-planar features which 
are found to be located both behind the disk and in front of the disk, could be in the plane 
of the LMC disk itself if there were an over-estimate or under-estimate 
of the reddening respectively. 
It has been demonstrated by \cite{Z97} that the 
extinction property of the LMC varies both spatially and as a function 
of stellar population. In our study, the dereddening of RC stars is done 
using the reddening values estimated from the RC stars itself.
In order to understand the effect of reddening, we plotted a two dimensional plot of reddening as well 
as the deviations. The deviations and the reddening values obtained by applying method 1 as well as method 2 
are given in Fig 13. The lower left and lower right panels show the reddening 
distribution obtained from method 1 and method 2 respectively. The upper left and upper right panels show 
the distribution of the deviations obtained from method 1 and method 2 respectively. 
In the lower panels, the size of the point is proportional to the reddening value and in the
upper panels the size of the point is proportional to the amplitude of the deviation. The red points
in the upper panels of the plots represent the regions behind the plane and blue points represent
the regions in front of the plane. 
The
regions in the southwestern part of the LMC disk around our suggested warps show more reddening.
Thus if we assume a constant reddening across the LMC we will not be able to identify these warps. At
the same time, we do not see a strong correlation between reddening and the deviation, because
both positive and negative deviations are observed for regions with large reddening. That is, the
reddening could not have been both under and over-estimated.

\begin{figure}
\resizebox{\hsize}{!}{\includegraphics{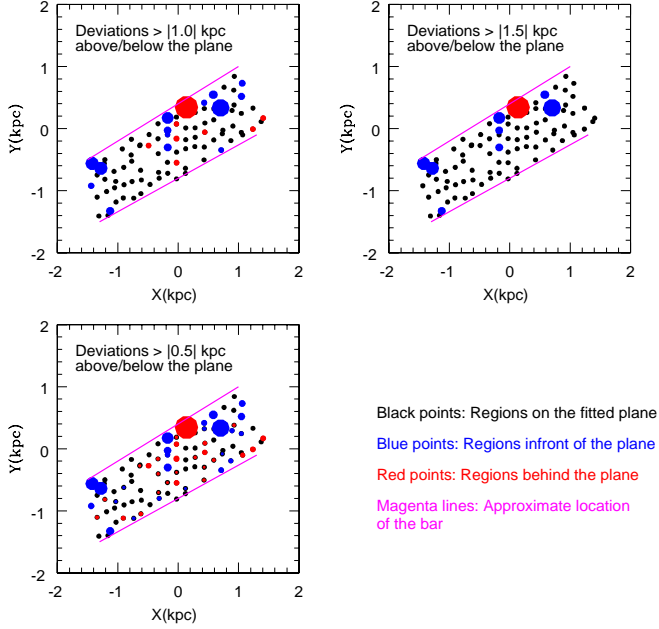}}
\caption{The regions in the bar of the LMC. The size of the red and blue points is proportional to 
the amplitude of the deviation.}   
\end{figure}

\section{Comparison with previous estimates}
Previously many studies have been done to obtain the planar parameters of the LMC disk using various 
tracers. The values obtained from those studies along with our estimates are summarised in Table 1. 
Tracers used in those studies are also mentioned in the table. From Table 1 we can see that the 
structural parameters have a range of values. The inclination value varies from 23$^\circ$$\pm$0.8$^\circ$ 
to 37$^o$.4$\pm$2$^o$.3 and PA$_{lon}$ varies from 122$^o$.5$\pm$8$^o$.3 to 170$^o$$\pm$5. Our estimates 
are found to be with in this range. \cite{SS10} 
suggested that the complicated structure of the inner LMC causes variation in the estimated planar 
parameters depending on the area covered for each study. In the present study also, in section 5 we 
showed that the inner structure of the LMC is disturbed with structural variations. 
Various studies of the LMC disk and bar regions (Fig.6 given in \cite{k09}, Fig.2 in \cite{ss09apj} and 
Fig.4 in \cite{s03}) have shown that it is a highly structured galaxy. Thus the variations in the estimate 
of the planar parameters are likely to be due to these structures.
\begin{table*}
\centering
\caption{Summary of orientation measurements of LMC disk plane}
\label{Table:1}
\vspace{0.25cm}
\begin{tabular}{lrrrr}
\hline \\
Reference & Inclination, $i$  & PA$_{lon}$, $\phi$  
& Tracer used for the estimate\\ \\
\hline
\hline \\

\cite{df72} & 27$^o$$\pm$2$^o$ & 170$^o$$\pm$5 & Isophotes \\
\cite{F77} & 33$^o$.0$\pm$3$^o$ & 168$^o$$\pm$4$^o$ & HI\\
\cite{cc86} & 28$^o$.0$\pm$5$^o$.9 & 142$^o$.4 $\pm$7$^o$.7 & Cepheids\\ 
\cite{LR92} & $-$ & 162$^o$.0 & HI\\
\cite{k98} & 22$^o$.0$\pm$6$^o$ & 168$^o$.0 & HI \\
\cite{vc01} & 34$^o$.7$\pm$6$^o$.2 & 122$^o$.5$\pm$8$^o$.3 & AGB stars \\ 
\cite{os02} & 35$^o$.8$\pm$2$^o$.4 &145$^o$$\pm$4$^o$ & Red clump stars \\
\cite{n04} & 30$^o$.7$\pm$1$^o$.1 & 151$^o$$\pm$2$^o$.4 & Cepheids\\
\cite{p04} & 27$^o$.0$\pm$6$^o$.0 & 127$^o$$\pm$10$^o$.0 & Cepheids\\
\cite{k09} & 23$^o$.5$\pm$0$^o$.4 & 154$^o$.6$\pm$1$^o$.2 & Red clump stars \\ 
\cite{SS10} (OGLE III data)& 23$^o$$\pm$0$^o$.8 & 163$^o$.7$\pm$1$^o$.5 & Red clump stars\\
\cite{SS10} (MCPS data) & 37$^o$.4$\pm$2$^o$.3 & 141$^o$.2$\pm$3$^o$.7 & Red clump stars\\\\
\hline
\hline \\
Our estimates\\ 
\hline \\
Method1 (H band magnitudes)&26$^o$.6 $\pm$ 1$^o$.3 & 148$^o$.3$\pm$3$^o$.8& Red Clump stars\\
Method2 (H band magnitudes)&25$^o$.7$\pm$1$^o$.6& 141$^o$.5$\pm$4$^o$.5& Red Clump stars\\
Method2 (J band magnitudes)&24$^o$.4$\pm$2$^o$.5 & 146$^o$.3$\pm$6$^o$.5& Red Clump stars\\
\hline\\

\hline
\end{tabular}
\end{table*}

\begin{figure}
\resizebox{\hsize}{!}{\includegraphics{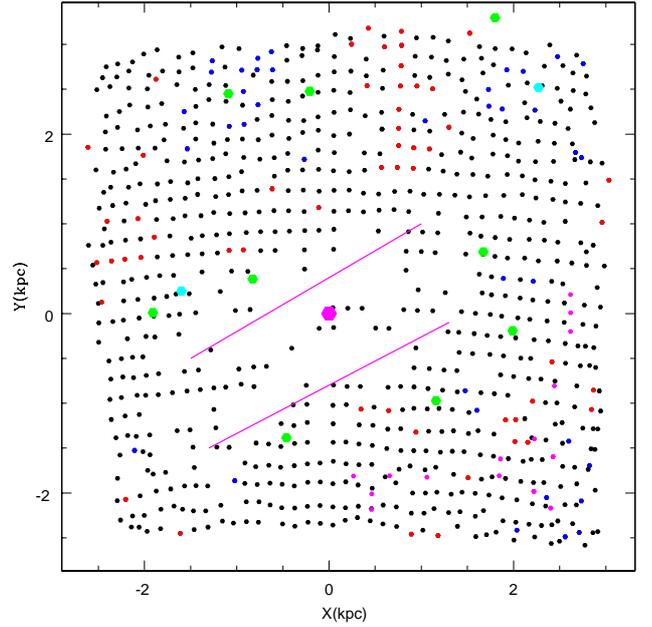}}
\caption{The regions used in the analysis based on method 1 are shown. The black points are regions on 
te fitted plane. The blue and red points are the regions which are 
infront of the plane and behind the plane respectively. The green points are the 
locations super giant shells given by Book et al. (2008). The cyan points are the location of major 
star forming regions 30 Dor and N11. The magenta lines are the approximate location of the bar. The magenta  
hexagon is the optical center of the LMC.}   
\end{figure}

\begin{figure}
\resizebox{\hsize}{!}{\includegraphics{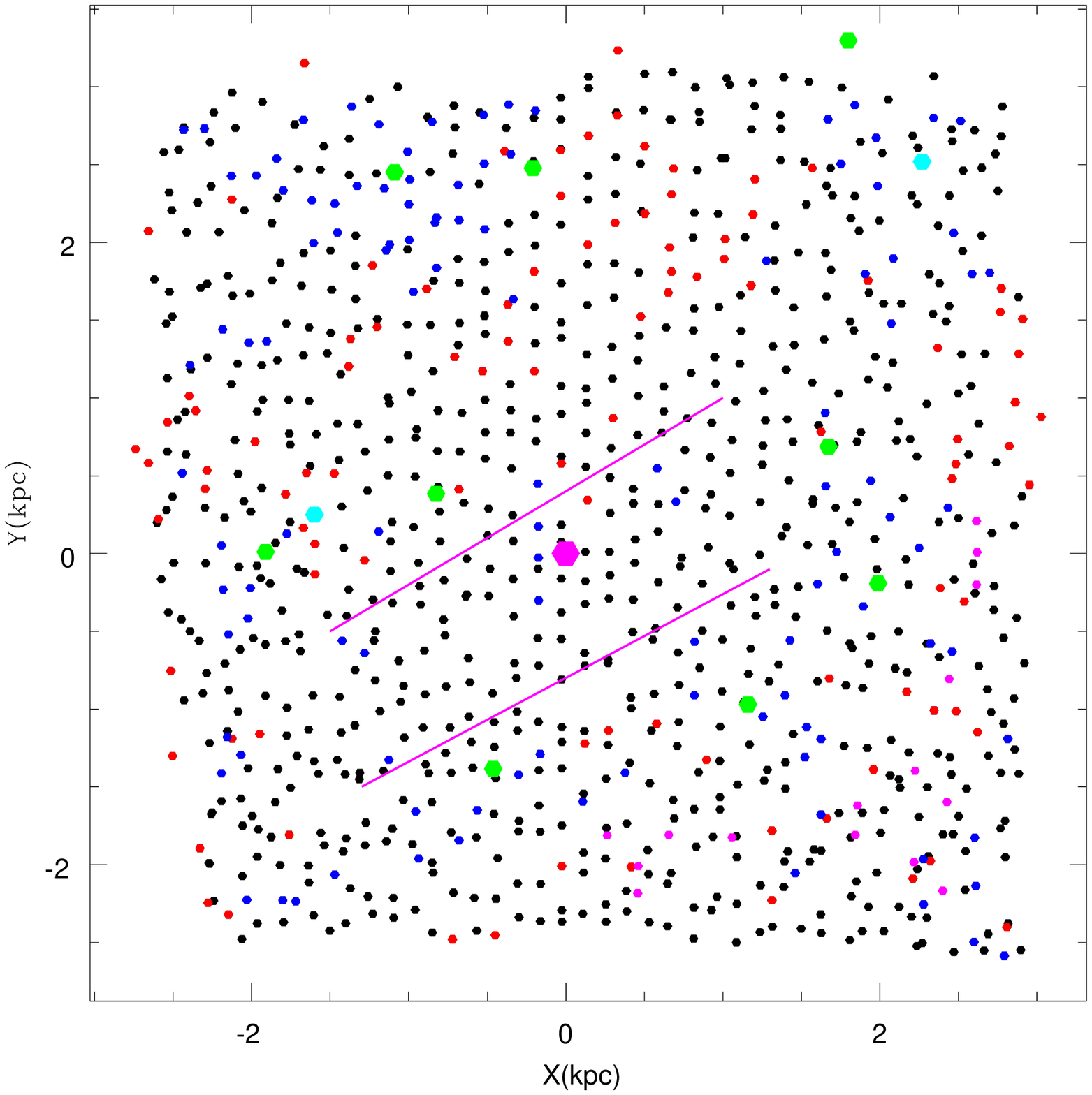}}
\caption{The regions used in the analysis based on method 2 are shown. The colour coding and the symbols 
are the same as in Fig 15.}   
\end{figure}

\subsection{Comparison with the study of K09} 
In our present study we used the same data set and tracer used by K09 to understand the structure 
of the LMC disk. In this section we compare the results of their study with ours in detail. K09 obtained an 
inclination of 23$^o$.5 $\pm$ 0$^o$.4 and PA$_{lon}$ to be 154$^o$.6 $\pm$ 1$^o$.2 for the LMC disk. Our estimates 
of the inclination is slightly higher than the value obtained by K09 and we estimate a lower value for PA$_{lon}$. 
The variation in the values of PA$_{lon}$ and inclination may be due the differences in the methodologies 
adopted in both the analysis. There is an overlap of sub-regions in the edges of the data set in the analysis 
of K09. In our analysis there is no overlap of sub-regions. The contamination of the RC sample by the stars 
in other evolutionary phases is less in our study compared to the study of K09. Also, the reddening correction 
adopted by K09 is not very 
clear. If K09 adopted a reddening correction on a star-by-star basis, the intrinsic colour chosen by 
K09 is for the RC stars. The reddening correction on a star by star basis is applicable only if all the stars 
in the sample are RC stars. The sample of K09 has AGB stars and also stars in other evolutionary phases. 
A method less susceptible to contamination is to deredden the observed peak colour corresponding to the 
RC stars in each sub-region. 
Again, from Fig 1 of K09 which shows the location of selected sample we can see that there are stars 
which have (J$-$H) colour less than 0.47 mag (0.47 is the intrinsic colour chosen by K09). Though we 
expect a spread in the colour (RC stars which are slightly bluer and slightly redder than the intrinsic colour) 
of the RC distribution due to population effects, internal reddening and photometric errors, the colour selection 
range used by K09 is very large and includes stars from other evolutionary phases.  If K09 has adopted 
star by star reddening correction, then it is not clear how they did correction for those stars which are bluer 
than 0.47 mag. The above mentioned differences could have resulted in the variation of the estimated parameters 
in the two studies.

K09 found that the bar floats above the disk by 1 kpc and is closer to us.  The bar regions are included in our 
analysis based on method 2. In our study most of the bar regions are on the plane of the LMC disk. 
As the error associated 
in the estimation of the distance is $\sim$ 1.4 kpc, regions which show deviations less than 1.5 kpc (1 sigma) 
are not considered as the real deviations in our analysis. In that case if the bar is brighter than the disk 
by only 1 kpc then it is not identified as a feature closer to us. In order to clarify this, we plotted the 
regions in the bar which show deviation greater than 0.5 kpc along with those which are on the plane. 
The plot is shown in the lower left panel of Fig. 14. The upper left and upper right panels show regions 
which deviate greater than 1 kpc and 1.5 kpc respectively along with the regions on the plane.
In all the panels, the blue points are those which are in front of the plane, the red are the regions which are 
behind the plane and the black points are those which are on the plane.  From the lower panel of the plot we 
can see that in the bar all three types of regions are present. Among the deviated regions there are regions 
which are nearer to us as well as away from us. These suggest that the bar is part of the disk and is structured. The middle and the upper panel plots 
also suggest the same. As mentioned in the earlier section, the differences in the methodologies adopted in 
our study and that of the K09 may be the reasons for the difference in the results. 

K09 found that the major star forming region like 30 Dor ($\sim$ at X=-1.6, Y=0.25) and N11 ($\sim$ at X=2.27, 
Y=2.52) are behind and in front of the plane respectively. The super giant shell (given in Book et al. 2008), 
LMC4 ($\sim$ at X=-1.1, Y=2.45) is also found to be in front of the plane. 
In order to compare the locations of these deviations, we over plotted the deviations identified by 
K09 on the deviations identified in the present study based on method 1 and method 2 in Fig 15 and Fig 16 
respectively. The location of 30 Dor and N11 are shown as cyan points. In the figure we have also shown the 
location of super giant shells given by Book et al (2008) as green points. In the upper panel, the 30 Dor 
region lies exactly on top of behind the plane features. In the lower panel, the 
30 Dor region is closer to the behind the plane feature. Thus we also find that the 30 Dor 
region is probably behind the disk. The star forming region, N11 also lies closer to the features located 
infront of the disk. The super giant shell LMC4 ($\sim$ at X=-1.1, Y=2.45) also lies nearer to us. These 
results match in both the studies. 

The results of K09 did not reflect the warps identified in the south - western regions identified by 
\cite{os02} and also by \cite{SS10} from the optical studies. Our present study identifies the warps 
in the south western region. K09 finds in this region a deviation 
from the plane in the opposite direction which is not very significant. In our analysis also, we find some small 
amplitude deviations away from the plane in the south-western region. But along with these deviations we find 
some regions which are significantly deviated from the plane in such a way that they are closer to us. 
The non identification of warps in the study of K09 may be due to two factors. One may be the lack of 
appropriate reddening correction by K09 for the RC stars in the south western region. From Fig. 6 and Fig. 7
we can see that the reddening is relatively high in the south western regions. A large reddening was also 
found in the south western region by \cite{os02}. The second factor which may have 
contributed is the overlapping of sub-regions. In the analysis of K09 there is an overlapping of 
sub-regions in the south western end. From Fig 8 
(upper panels) we can see that there are both in front of the plane feature as well as the behind the plane 
feature which may get averaged out if the regions are overlapped.

\section{Conclusions}

$\bullet$ The structure of the LMC disk is studied using the J and H magnitudes (near IR magnitudes) of the 
RC stars. We find an inclination of 26$^o$.6 $\pm$ 1$^o$.3 \& PA$_{lon}$ = 148$^o$.3$\pm$3$^o$.8 
for the disk when the peaks were identified by profile fitting. When the peaks of the red clump 
distribution were identified by averaging, an inclination of  25$^o$.7$\pm$1$^o$.6 and 
PA$_{lon}$ = 141$^o$.5$\pm$4$^o$.5 were obtained.\\

$\bullet$ A reddening map based on $(J-H)$ colour of RC stars is presented.\\ 

$\bullet$ We estimate a distance modulus, $\mu$=18.48$\pm$0.1 mag to the LMC.\\

$\bullet$ The bar is found to be part of the disk, within 500 pc.\\

$\bullet$The inner (r$<$ 3$^o$.0) and outer (r$>$ 3$^o$) disk structures of the LMC are 
found to be different. \\

$\bullet$ We identified extra-planar features in the disk similar to those identified 
from the optical analysis of the RC stars. \\

$\bullet$ From the comparison of our results with the study of K09, who used the same data set and tracer, 
we found that the selection of the RC sample from optical 
identification helps to reduce the contamination of the sample by stars in other evolutionary phases.\\

\end{document}